\documentclass[11pt,twoside]{article}
\usepackage{asp2004}
\usepackage{psfig}
\usepackage{epsf}
\usepackage{graphics}
\usepackage{lscape}
\def\edcomment#1{\iffalse\marginpar{\raggedright\sl#1\/}\else\relax\fi}
\reversemarginpar

\begin{document}
\title{Why there are no jets from cataclysmic variable stars}
\author{Jean-Pierre Lasota$^1$, Noam Soker$^2$}
\affil{$^1$Institut d'Astrophysique de Paris, 98bis boulevard
Arago, 75014 Paris, France; lasota@iap.fr}
\affil{$^2$Department
of Physics, Technion - Israel Institute of Technology, Haifa
32000, Israel; soker@physics.technion.ac.il}

\begin{abstract}
We argue that the recent thermal model of jet launching by young
stellar objects, when applied to system containing disk-accreting
white dwarfs naturally explain the otherwise astonishing absence
of jets in cataclysmic variable stars. Thermal launching is
possible when the accreted material is strongly shocked due to
large gradients of physical quantities in the boundary layer (or
at the inner boundary of a truncated disk) and then cools on a
time scale longer than its ejection time from the disk. In our
framework the magnetic fields are weak, and serve only to
recollimate the outflow at large distances from the source, or to
initiate the shock, but not as a jet-driving agent. Using criteria
for shock formation and mass ejection, we find the mass accretion
rate above which jets can be launched from boundary layers around
accreting white dwarfs to be $\dot M_{\rm WD} \ga 10^{-6}
\textrm{M}_\odot \textrm{y}^{-1}$, which explains the absence of
jets in cataclysmic variable stars and their presence in other
white-dwarf accreting systems such as super-soft X-ray sources,
symbiotic stars and classical novae.
\end{abstract}
\thispagestyle{plain}

\section{Introduction}

It is widely believed that accretion with angular momentum leads
to ejection of jets. This belief is based on observations of jets
in Young Stellar Object (YSOs), Low-Mass and High-Mass X-ray
Binaries (LMXBs \& HMXBs) and Active Galactic Nuclei. Cataclysmic
Variables (CVs), however, are a blatant exception from the
presumed universality of the accretion -- jet connection. However,
CVs are not an exception from a more general, supposedly
universal, relation between accretion and ``ejection". Indeed,
although no jets have ever been observed in CVs, some of them emit
winds. For example, P Cygni profiles in resonant UV lines are
observed in some very luminous CVs such as the nova-like stars and
dwarf novae at outburst maximum. These winds are too cold to be
ejected by a thermal mechanism and are most probably driven by
radiative line pressure with some help of magnetic fields (see
e.g. Proga, these proceedings). But they are winds, even if
slightly collimated, but not jets.

The presence of a white dwarf in the center of an accretion flow
cannot be considered to be responsible for this jet-blowing
impotence since jets are observed in other accreting white-dwarf
systems such as Super Soft X-ray Source (SSXS), symbiotic stars
and novae. Fast, $\sim 1000-5000 \rm \ km\ s^{-1}$, collimated
outflows have been observed in some SSXSs, RX J0513.9-6951, RX
J0019.8+2156 and RX J0925.7-4758 \citep[see references
in][hereafter SL04]{SL04} and symbiotic systems are also known to
blow jets \citep[][and references therein]{sok1,Bro}. Also the
fast nova V1494 Aql produced high velocity $\sim 2800 \rm \ km\
s^{-1}$ jets during its decline from outburst maximum
\citep{Iijima,retter04}. SSXSs, symbiotic systems and classical
novae differ from CVs by much higher accretion rates and the
presence of an additional source of energy produced by
thermonuclear reactions. Although novae are CVs undergoing a
thermonuclear runaway for the purpose of the present investigation
we will separate them from the other members of this class of
binaries. \citet{Livio00} speculated that this latter difference
might account for lack of jets in CVs, however, without providing
a physical explanation.

Recently we (SL04) have shown that this difference is naturally
explained if jets in accreting white-dwarf systems are produced by
a thermal mechanism. As discussed below, magnetic fields would
play a role in jet formation but this would be rather auxiliary,
the main thrust being provided by thermal pressure.

\section{Too small discs?}

First we will consider another suggestion by \citet{Livio00}
according to whom the absence of jets in CVs might be related to
the small size of their accretion discs. The argument is based on
numerical jet models of poloidal collimation \citep[see
e.g.][]{ol98} in which the vertical component of the magnetic
field varies as
\begin{equation}
B_z \sim r^{-1}
\label{poloidal}
\end{equation}
and the jet opening angle is
\begin{equation}
\theta\sim \left(\frac{R_{\rm in}}{R_{\rm out}}\right)^{1/2}.
\label{angle}
\end{equation}
When discs extend down to the smallest possible circular orbit, CV
discs are much less extended that those of LMXBs simply because a
$\sim 10 \textrm{M}_{\sun}$ black hole is much smaller than a
white dwarf. However, observations and models suggest that LMXB
discs are often truncated, in particular during jet launching
\citep*[see e.g.][]{fbg}. In this case the two types of systems
might have similar disc sizes.

Indeed, taking for the outer disc radius $R_{\rm out}=0.9 R_{\rm
L_{1}}$, where  $R_{\rm L_{1}}$ is the {\sl mean} Roche-lobe
radius \citep[see e.g.][for the formula]{apa02}, one obtains for
the ratio of the CV to LMXB putative jet opening angles
\begin{equation}
\frac{\theta_{\rm{CV}}}{\theta_{\rm{BH}}}\approx 7.5
\left(\frac{M_{\rm{BH}}}{M_{\rm{WD}}}\right)^{1/3}
\left(\frac{M_{\rm{BH}}}{10\ \rm{M}_{\odot}}\right)^{-1/2}
\left(\frac{R_{\rm WD}}{5 \times 10^8 \rm cm }\right)^{1/2}
\left(\frac{R_{\rm in}}{3 R_{\rm S}}\right)^{-1/2}
\label{angles}
\end{equation}
where $R_{\rm S}=2GM/c^2$ is the Schwarzschild radius and we
assumed a 1 $\textrm{M}_\odot$ white dwarf. Since jets are
observed during hard/low states when (according to a popular
scenario) $R_{\rm in} \sim 100 R_{\rm S}$, the extent of the disc
of a jet-launching LMXB could be comparable to that of a typical
CV which makes the small disc argument not very compelling. One
should add that although CVs discs can also be truncated, one
expects jet to be launched at high accretion rates at which the
accretion disc would reach down to the white dwarf surface.

\section{Thermal launching}

In jet models magnetic fields may play a dominant role in three
types of processes: in triggering the \emph{jet ejection events},
e.g., by causing instabilities in the disk, in accelerating the
jets \citep[as in the classic ``centrifugal wind'' mechanism,
first proposed by][]{BP82} and finally in collimating the jets
\citep[e.g.][]{HN89}. Since models containing all these three
elements fail to account for the absence of jets in CVs one is
justified in trying to replace at least one of them by a different
mechanism. SL04 showed that depriving the magnetic field of its
accelerating role and replacing it by the action of the thermal
pressure not only offers a natural explanation of the absence of
jets in CVs but also accounts for the presence of jets in other
white-dwarf systems such as classical novae, SSXSs and symbiotic
stars.

SL04 based their argument on the model of thermal pressure
acceleration proposed by \citet{Tor1} and \citet{TorG} recently
developed and extended  by \citet{SR03} to explain strongly
collimated outflows in YSOs. In this jet model  magnetic fields
are weak, and might serve only to re-collimate the outflow at
large distances from the source and might trigger disturbances in
the boundary layer (BL) where the disk adjust itself to the
conditions at the surface of the accreting star.

\citet{SR03} found two conditions necessary for the jet thermal
launching model to work. The first \emph{condition is} that the
strongly shocked gas in the BL cools slowly so that the thermal
pressure have enough time to accelerate the jet's material. The
second condition requires that weakly shocked blobs in the BL
expand and disturb it in such a way that a strong shock develops.
SR03 term such strong shocks `spatiotemporally localized (but not
too small!) accretion shocks', or SPLASHes.

\subsection{Ejection condition}

The characteristic radiative cooling time in the BL is equal to
the photon diffusion time
\begin{equation}
t_{\rm cool} = H^2\, \frac{\rho \kappa}{c} ,
\label{taudiff1}
\end{equation}
where $H$ is the vertical scale height, $\rho$ the density,
$\kappa$ the opacity, and $c$ the speed of light, whereas the
ejection time is given by the dynamical time
\begin{equation}
t_{\rm eject} = \frac{H}{\sqrt{2} v_{\rm K}},
\label{tauesc}
\end{equation}
where $v_{\rm K}=\sqrt{GM_1/R}$, where $R$ is the distance to the
disc's center.

Using the the mass conservation equation $\dot M = 2 \pi R 2H \rho
v_r$, taking for the radial velocity $v_r \simeq \alpha
\left(H/R\right)^2 v_{\rm K}$ and and taking into account the
strong shock density-contrast condition one obtains
\begin{equation}
\frac {t_{\rm cool}}{t_{\rm ej}} \simeq \nonumber
\frac{\dot M \kappa}{\pi c \alpha  \epsilon^2 R} \\
\simeq 1.3 \left( \frac{H/R}{0.1} \right)^{-2}
\left(\frac{\alpha}{0.1} \right)^{-1} \left(\frac{\dot M}{10^{-7}
\rm M_\odot  y^{-1}}\right) \left(\frac{\kappa(\rho,T)}{\rm cm^2
g^{-1}} \right) \left(\frac{R} {\rm R_{\sun}} \right)^{-1}.
\label{taudiff3}
\end{equation}
Therefore for accreting white dwarfs ($R\simeq 0.01 R_\odot$) the
condition $t_{\rm cool} \ga t_{\rm ej}$ is satisfied for
\begin{equation}
\dot M_{s} \ga 2 \times 10^{-9} \left( \frac{H/R}{0.1} \right)^{2}
\left( \frac{\alpha}{0.1} \right) M_\odot \rm yr^{-1},
\label{accwd1}
\end{equation}
which is satisfied for the nova-like stars and dwarf-novae at
maximum, i.e. for CVs at highest accretion rate. However, in these
systems one observes only winds, not jets.

\subsection{Strong shock condition}

The model assumes that hundreds of small blobs are formed in the
sheared BL \citep[section 2 of][]{SR03}. The blobs occasionally
collide with each other, and create shocks which cause the shocked
regions to expand in all directions. If the shocked regions
continue to expand out into the path of yet more circulating
blobs, stronger shocks may be created, as was proposed by
\citet{PS79} to explain the emission of X-rays out of disk BLs in
dwarf novae. For the shocked blobs to expand, the radiative
cooling time of {\em individual blobs}, $t_{\rm cool} \simeq
\ell^2 \kappa \rho_b/c$, must be longer than their adiabatic
expansion time $t_{\rm ad} = \ell/c_s$, where $\ell$ is the size
of an expanding blob, and $\rho_b$ the post-shock blob's density.

This condition also leads to a minimum value for the mass
accretion rate \citep[][eq. 12 in SL04]{SR03}
\begin{equation}
\dot M_b \ga 4.2 \times 10^{-5} \frac{1}{\kappa(\rho_b,T_b)}
\left( \frac{\alpha}{0.1} \right) \left(\frac{R_j}{\rm R_{\sun}}
\right) \rm M_\odot y^{-1},
 \label{acc01}
\end{equation}
where $R_j$ is the radius from where the jet is launched. For
accreting white dwarfs the weak-shock temperature $T_b \ga 5
\times 10^6$ K. Therefore also in this case $\kappa=0.4 $cm$^2$
g$^{-1}$ (SL04). The strong-shock formation condition for
accreting white dwarfs is
\begin{equation}
\dot M_{\rm WD} \ga 10^{-6} \left( \frac{\alpha}{0.1} \right) \rm
M_\odot \rm yr^{-1}. \label{accwd2}
\end{equation}
This is roughly two to three orders of magnitude larger than the
\emph{maximal} accretion rate in CVs (the accretion rate of
nova-like stars and dwarf-novae at maximum is always $\la
10^{-8}\rm M_\odot \rm yr^{-1}$). The condition (\ref{accwd2})
provides therefore an explanation for the absence of jets in CVs.

As mentioned in the introduction this explanation is strengthened
by the fact that white-dwarf systems accreting at rates satisfying
Eq.(\ref{accwd2}) \emph{do show} jets. In SSXSs white dwarfs
accrete at rates of $3 \times 10^{-8}-10^{-6}\ \rm M_\odot
yr^{-1}$ from a companion, and sustain nuclear burning on their
surface \citep[e.g.][]{vandh}. In symbiotic systems, white dwarfs
accrete at high rates from the wind of red giant branch stars or
asymptotic giant branch stars. In some of the symbiotic systems
which blow jets the white dwarf sustains a quasi-steady nuclear
burning, similar to SSXSs; in others, there is no nuclear burning
\citep{Bro}.

\citet{retter04} showed that conditions during the jet ejection by
V1494 Aql are consistent with condition Eq. (\ref{accwd2}). He
estimated the accretion rate  to be then $\sim 10^{-6}\rm\ M_\odot
\rm yr^{-1}$. To be really consistent with our model, the white
dwarf should be accreting from a disc during jet production. The
disc is most probably destroyed during the nova explosion and
reforms during the decline. In V1494 Aql the accretion disc would
have to be present during the ``transition phase", three months
after the maximum. This is rather rapid compared to another fast
nova GQ Mus in which the disc reappeared only after a decade when
the X-ray source had turned off \citep{staridan}, but still
another fast nova V1974 Cyg might have shown a disc signature some
thirty months after the outburst \citep{retter97}.

\section{Discussion and perspectives}

The thermal model elucidates why there exists a critical accretion
rate necessary for jet launching. This critical rate is well above
the maximum rate encountered in CVs and so solves the mystery of
their jet quietness. One could argue that this is just a
coincidence. The critical value could, for example, correspond to
the appearance of a large-scale (poloidal) magnetic field
necessary for a \citeauthor{BP82} -- type mechanism to work, as
proposed by \citet{Livio03}. According to these authors, at a
critical rate the accretion disc would switch from a standard
radiative disc to a state where most of the accretion energy is
released in the form of a bulk flow. For the time being no MHD
simulation is capable to follow such a process (see the article by
Balbus in these proceedings), and the value of the critical
accretion rate (if any) can be only matter of speculation.
\citet{Livio03} suggested that the inner parts of the nova-like
star discs are underluminous because of the transition into an
outflow phase. This would imply a critical accretion rate of $\sim
10^{-8}\rm M_\odot \rm yr^{-1}$. However, this rate corresponds
rather to launching of winds, not of a jets, and the mystery of
their absence in CVs would be still with us. It should be also
noted that the alleged luminosity deficit in the inner disc of
nova-like stars could be (at least in part) just an artefact of
the disc model used \citep{Smak94}. In any case the existence of a
critical accretion rate for the presence of large-scale poloidal
field would not be necessarily in contradiction with our criterion
Eq. (\ref{accwd2}) which is only a necessary condition for jet
launching. On the other hand, if the poloidal field is generated
at the cost of local energy dissipation thermal jet launching
could be problematic.

Finally, our model has the vocation to be universal despite the
fact that we use properties of the boundary layer which would not
exist when accretion occurs onto black holes or strongly
magnetized stars. However, the boundary layer is important only
because it is where strong shocks can be produced. The required
strong gradients could be also produced in accretion discs by
``magnetospheric" MHD even when the central object is a black hole
\citep[e.g.][and references therein]{Linar}. In such a case MHD
instabilities, turbulence, or other disturbances may lead to
strong shocks; the high post-shock pressure may accelerate gas and
form jets and/or winds, e.g., as was shown for non-radiative
accretion around a black hole by \citet{Devil03}.

Scaled to the case of an accreting black hole Eq. ({\ref{acc01}})
becomes
\begin{equation}
\dot m \equiv \frac{\dot M}{\dot M_{\rm Edd}} \ga 0.02 \left(
\frac{\alpha}{0.1} \right) \left( \frac{0.4 \rm \ g \ cm^{-2}}
{\kappa}\right) \frac{R_j}{R_{\rm G}},
\label{comp}
\end{equation}
where $\dot M_{\rm Edd}= L_{\rm Edd}/0.1c^2= 2.3\times 10^{-8}
\left(M/{\rm M}_{\odot}\right) \rm M_\odot y^{-1}$ is the
Eddington accretion rate. In \citet{SL04} we mistakenly claimed
that this formula might be relevant to the appearance of steady
jets (we Friedrich Meyer for pointing this out at the present
conference). It remains to be seen how and when the thermal
jet-launching model applies to systems with black hole, but we
expect \citetext{in preparation} that in the case of microquasars
it would be rather relevant to the launching of powerful, high
Lorentz-factor jets \citep[e.g.][]{fbg}.


\begin{thebibliography}{}


\bibitem[Blandford, \& Payne(1982)]{BP82} Blandford, R.~D.~\&
Payne, D.~G.\ 1982, \mnras, 199, 883

\bibitem[Brocksopp, et al.(2004)]{Bro} Brocksopp, C., Sokoloski, J. L.,
Kaiser, C., Richards, A. M., Muxlow, T. W. B., Seymour, N. 2004,
MNRAS, 347, 430

\bibitem[van den Heuvel, et al.(1992)]{vandh} van den Heuvel, E. P. J.,
Bhattacharya, D., Nomoto, K., \&  Rappaport, S. A. 1992, A\&A,
262, 97

\bibitem[De Villiers, Hawley, \& Krolik(2003)]{Devil03} De
Villiers, J., Hawley, J.~F., \& Krolik, J.~H.\ 2003, \apj, 599,
1238

\bibitem[Fender, et al.(2004)Fender, Belloni, \& Gallo]{fbg} Fender,
R.~P., Belloni, T.M., \& Gallo, E.\ 2004, \mnras, in press

\bibitem[Frank, King, \& Raine(2002)]{apa02} Frank, J., King,
A., \& Raine, D.~J.\ 2002, Accretion Power in Astrophysics: Third
Edition, by Juhan Frank, Andrew King, and Derek
J.~Raine.~Cambridge University Press, 2002, 398 pp.

\bibitem[Heyvaerts, \& Norman(1989)]{HN89} Heyvaerts, J.~\&
Norman, C.\ 1989, \apj, 347, 1055

\bibitem[Iijima, \& Esenoglu(2003)]{Iijima} Iijima, T.~\&
Esenoglu, H.~H.\ 2003, \aap, 404, 997

\bibitem[Li \& Narayan(2004)]{Linar} Li, L.~\& Narayan, R.\
2004, \apj, 601, 414

\bibitem[Livio,(2000)]{Livio00} Livio, M. 2000, in Cosmic Explosions, Eds.
 S.S. Holt \& W.W. Zhang, (American Institute of Physics Conf.) 522, 275

\bibitem[Livio, Pringle, \& King(2003)]{Livio03} Livio, M.,
Pringle, J.~E., \& King, A.~R.\ 2003, \apj, 593, 184

\bibitem[Ogilvie, \& Livio(1998)]{ol98}Ogilvie, G.I. \& Livio, M. 1998, \apj, 499, 329).

\bibitem[Pringle, \& Savonije(1979)]{PS79} Pringle, J.~E.~\&
Savonije, G.~J.\ 1979, \mnras, 187, 777

\bibitem[Retter,(2004)]{retter04}Retter, A. 2004, \apjl, in press
(astro-ph/0408553)

\bibitem[Retter, Leibowitz, \& Ofek(1997)]{retter97} Retter, A.,
Leibowitz, E.~M., \& Ofek, E.~O.\ 1997, \mnras, 286, 745

\bibitem[Smak(1994)]{Smak94} Smak, J.\ 1994, Acta Astronomica,
44, 265

\bibitem[Soker, \& Lasota(2004)]{SL04} Soker, N. \& Lasota, J.-P.
2004, \aap, 422, 1039 (SL04)

\bibitem[Soker, \& Regev(2003)]{SR03} Soker, N.~\& Regev, O.\
2003, \aap, 406, 603

\bibitem[Sokoloski, et al.(2004)]{sok1} Sokoloski, J.L., Kenyon,
S.L., Brocksopp, C., Kaiser, C.R. \& Kellogg, E.M. 2004,
RevMexAA(SC), 20, 35

\bibitem[Starrfield, et al.(1994)]{staridan} Starrfield, S.,
Idan, I., Shaviv, G., Shore, S.~N., Krautter, J., \& Sonneborn,
G.\ 1994, Bulletin of the American Astronomical Society, 26, 1324

\bibitem[Torbett,(1984)]{Tor1} Torbett, M. V. 1984, ApJ, 278, 318

\bibitem[Torbett, \& Gilden(1992)]{TorG} Torbett, M. V., \& Gilden, D. L.
1992, A\&A, 256, 686


\end{thebibliography}
\end{document}